\documentclass[conference]{IEEEtran}
\usepackage{color}
\usepackage{enumerate}
\usepackage[cmex10]{amsmath} 
\usepackage{siunitx}
\usepackage{import} 
\usepackage{enumitem}

\usepackage{indentfirst} 
\usepackage{standalone}
\usepackage[section]{placeins} 
\usepackage{pdflscape}
\usepackage{tabularx}
\usepackage{algorithm} 
\usepackage{amssymb} 
\usepackage{microtype} 
\usepackage{soul} 
\usepackage{mathtools}
\usepackage{multirow}
\usepackage{verbatim}
\usepackage{graphicx}

\usepackage{pgfplots}

\usepackage{subfloat} 
\usepackage{verbatim}

\usepackage[utf8]{inputenc}
\usepackage[T1]{fontenc}
\usepackage{amsmath} 
\usepackage[shellescape,latex]{gmp} 
\usepackage[space]{grffile}
\usepackage{booktabs} 
\usepackage{xfrac}
\usepackage{tabularx}

\usepackage{cite}
\usepackage{adjustbox}

\usepackage{verbatim}
\usepackage{url}

\usetikzlibrary{shapes}

\usepackage{multicol, blindtext}    
\usepackage[]{color}
\usepackage{framed}
\usepackage{alltt}
\usepackage[T1]{fontenc}
\usepackage{array}
\usepackage{colortbl}
\usepackage{booktabs}
\usepackage{multirow}
\usepackage{eurosym}
\usepackage{listings}
\usepackage{ifthen}
\usepackage{pgfplots}
\usepackage{pgfplotstable}
\usepackage{pgfcalendar}
\usepackage{pgfgantt}
\usepackage{tikz}
\usetikzlibrary{shapes.geometric, arrows}

\usepackage{longtable}
\usepackage{amsfonts}
\usepackage{subfigure}
\usepackage{float}

\usepackage{mathrsfs}
\usepackage{algpseudocode} 
\usepackage{textcomp}
\usepackage{xspace}

\makeatletter
\newcommand\fs@betterruled{%
  \def\@fs@pre{\vspace*{5pt}\hrule height.8pt depth0pt \kern2pt}%
  \def\@fs@post{\kern2pt\hrule\relax}%
  \def\@fs@mid{\kern2pt\hrule\kern2pt}%
  \let\@fs@iftopcapt\iftrue}
\floatstyle{betterruled}
\restylefloat{algorithm}
\makeatother

\pgfplotsset{compat=1.16}

\usepackage{fmtcount} 

\newcommand{\NAME}{\mbox{\textsc{N}et\textsc{M}ind}\xspace}

\newcommand{\boldpar}[1]{\smallbreak\noindent\textit{#1.}}

\begin{document}
\title{\NAME: Adaptive RAN Baseband Function \\
\mbox{Placement by GCN Encoding and Maze-solving DRL}}

\author{\IEEEauthorblockN{
Haiyuan Li\IEEEauthorrefmark{1},
Peizheng Li\IEEEauthorrefmark{2},
Karcius Day Assis\IEEEauthorrefmark{3},
Adnan Aijaz\IEEEauthorrefmark{2},
Sen Shen\IEEEauthorrefmark{1},\\
Reza Nejabati\IEEEauthorrefmark{1},
Shuangyi Yan\IEEEauthorrefmark{1},
Dimitra Simeonidou\IEEEauthorrefmark{1}
}\\ 
\vspace{-0.3cm}
\IEEEauthorblockA{\IEEEauthorrefmark{1} Smart Internet Lab, University of Bristol, UK;
\IEEEauthorrefmark{2} Bristol Research and Innovation Laboratory, Toshiba Europe Ltd., UK;\\
\IEEEauthorrefmark{3} Federal University of Bahia, Brazil\\
Email: {\{ocean.h.li, sen.shen, reza.nejabati, shuangyi.yan, dimitra.simeonidou\}@bristol.ac.uk;}\\
{\{peizheng.li, adnan.aijaz\}@toshiba-bril.com;}
{karcius.assis@ufba.br}
}}

\maketitle

\begin{abstract}

The disaggregated and hierarchical architecture of advanced RAN presents significant challenges in efficiently placing baseband functions and user plane functions in conjunction with Multi-Access Edge Computing (MEC) to accommodate diverse 5G services. Therefore, this paper proposes a novel approach \NAME, which leverages Deep Reinforcement Learning (DRL) to determine the function placement strategies in RANs with diverse topologies, aiming at minimizing power consumption. 
\NAME formulates the function placement problem as a maze-solving task, enabling a Markov Decision Process with standardized action space scales across different networks. 
Additionally, a Graph Convolutional Network (GCN) based encoding mechanism is introduced, allowing features from different networks to be aggregated into a single RL agent. That facilitates the RL agent's generalization capability and minimizes the negative impact of retraining on power consumption.
In an example with three sub-networks, \NAME achieves comparable performance to traditional methods that require a dedicated DRL agent for each network, resulting in a 70\% reduction in training costs. Furthermore, it demonstrates a substantial 32.76\% improvement in power savings and a 41.67\% increase in service stability compared to benchmarks from the existing literature.

\end{abstract}


\vspace{-7pt}
\section{Introduction}
\label{sec:introduction}
5G networks, boasting capabilities such as enhanced Mobile Broadband (eMBB), massive Machine Type Communication (mMTC), and ultra Reliability and Low Latency Communication (uRLLC), have swiftly transitioned into widespread commercial deployment. In response to these advanced applications, the co-locating of Multi-access Edge Computing (MEC) and Radio Access Networks (RAN) have received much attention from academics and industries to establish rapid-response connections and reduce access latencies from users to their required services.

Aligning with the 3GPP standards~\cite{3gpp38801}, the state-of-the-art RAN differentiates itself from previous solutions and baseband partitioning standards that were once integrated into Remote Radio Unit (RRU) and Baseband Unit (BBU), and evolves into more disaggregated functions including Radio Unit (RU, wireless connectivity), Distributed Unit (DU, data layer processing), Core Unit User Plane (CU-UP, user data processing) and Core Unit Control Plane (CU-CP, control signaling mechanisms). These modules can be separated by a variety of baseband function splitting options tailored to different service needs. In addition, edge computing networks composed of MEC servers further empower RANs by carrying these RAN functions and bringing the User Plane Function (UPF) from the 5G Core (5GC) to the user side. The close proximity of RAN and UPF to users reduces the transmission distance between CU and UPF, making real-time data session routing more efficient. However, while this intricate module segmentation enhances the flexibility and granularity of network services, without meticulous management measures for the placement of baseband functions, it might lead to substantial resource redundancy and power inefficiencies.

Currently, two common approaches are used to address the baseband placement challenges, i.e., (Mixed) Integer Linear Programming ((M)ILP) and Deep Reinforcement Learning (DRL). ILP has been studied in works such as~\cite{bhamare2018efficient, rodriguez2020cloud, yu2020cu} for DU/CU placement in Centralized-RAN (C-RAN). Additionally, the authors in~\cite{harutyunyan2019latency, harutyunyan2020latency} explored optimal service function chain placement and resource allocation to reduce end-to-end latency. Meanwhile, study~\cite{zorello2022power} proposed using MILP to minimize DU/CU placement power consumption considering functional split, latency, and capacity constraints.
On the other hand, DRL approaches, as seen in research like~\cite{joda2022deep} explored DRL strategies for placing CUs and DUs and associating UEs to minimize end-to-end latency and Open RAN (O-RAN) deployment costs. Moreover, in~\cite{wang2022edge}, the use of graph neural networks in conjunction with DRL is investigated for optimizing DUs and CUs placement while provisioning lightpaths. Furthermore, ~\cite{mollahasani2021dynamic} delves into dynamic CU-DU selection using actor-critic learning in O-RAN for resource allocation.

However, it is noticed that current literature poses \textit{three critical drawbacks} in designing efficient baseband function placement policies for these advanced RAN architectures. 
\begin{itemize}[leftmargin=*]
    \item (M)ILP-based studies, while achieving near-optimal results, suffer from high computational complexity and time-consuming characteristics, rendering them unsuitable for latency-sensitive 5G services.
    \item Most previous DRL-based function placement literature described a non-Markovian environment where the future state is independent of the present. This suggests that the direct application of DRL to such environments may not be theoretically grounded.
    \item Current research primarily focuses on static network architectures, overlooking the challenges of model generalization ability in diverse network topologies and the consequent side effects of model retraining on power-saving performance.
\end{itemize}

To address the aforementioned issues, we proposed \NAME, a DRL paradigm enhanced with a Graph Convolutional Network (GCN) encoder for adaptive baseband placement\footnote{The realization of NetMind is detailed in the accompanying source code. Available at https://github.com/OCEAN-98/NetMind.}. The key functions and contributions of \NAME are summarized below:
\begin{itemize}[leftmargin=*]
    \item \NAME uniquely constructs the baseband placement in RAN as a \textit{maze-solving} problem, which can be framed as an MDP. A Deep Q Learning (DQN) algorithm is employed to minimize energy consumption by optimizing functional placement and routing provisioning.
    \item By incorporating \textit{GCN-based encoder and decoder} in \NAME, the state information from different networks can be unified into a consistent format and utilized to train a general DRL model. This model is universally \textit{applicable across networks} with varying structures, thereby, eliminating the retraining power costs in different scenarios.
    \item The training cost of \NAME is evaluated by comparing individual training for multiple networks with a single training session using combined state information from all networks. Results indicate that \NAME, trained once, needs only 30\% of the cumulative training costs.
    \item Comprehensive evaluation of \NAME performance is conducted. Taking the power consumption as metrics, the comparison of the proposal and three benchmarks, including (i) a random allocation solution~\cite{li2022energy} (ii) a greedy heuristic procedure designed by~\cite{casazza2017securing} and (iii) a MILP optimization proposed in~\cite{xiao2021energy}, is detailed.
\end{itemize}

The remainder of this paper is constructed as follows. 
Section~\ref{sec:problem formulation} presents the RAN scenario and formulates the optimization problem. Section~\ref{sec:Methodology} discusses the technique details of \NAME. Then, Section~\ref{sec:simulation} provides the simulation setup, the numerical results, and discussions regarding the \NAME.
Finally, the paper concludes with a summary of our key findings in Section~\ref{sec:conclusion}.

\section{Baseband Function Deployment Scenario in O-RAN and problem formulation}
\label{sec:problem formulation}
\begin{figure}[t]
    \centering
    \setlength{\abovecaptionskip}{0cm}
    \includegraphics[width=0.78\linewidth]{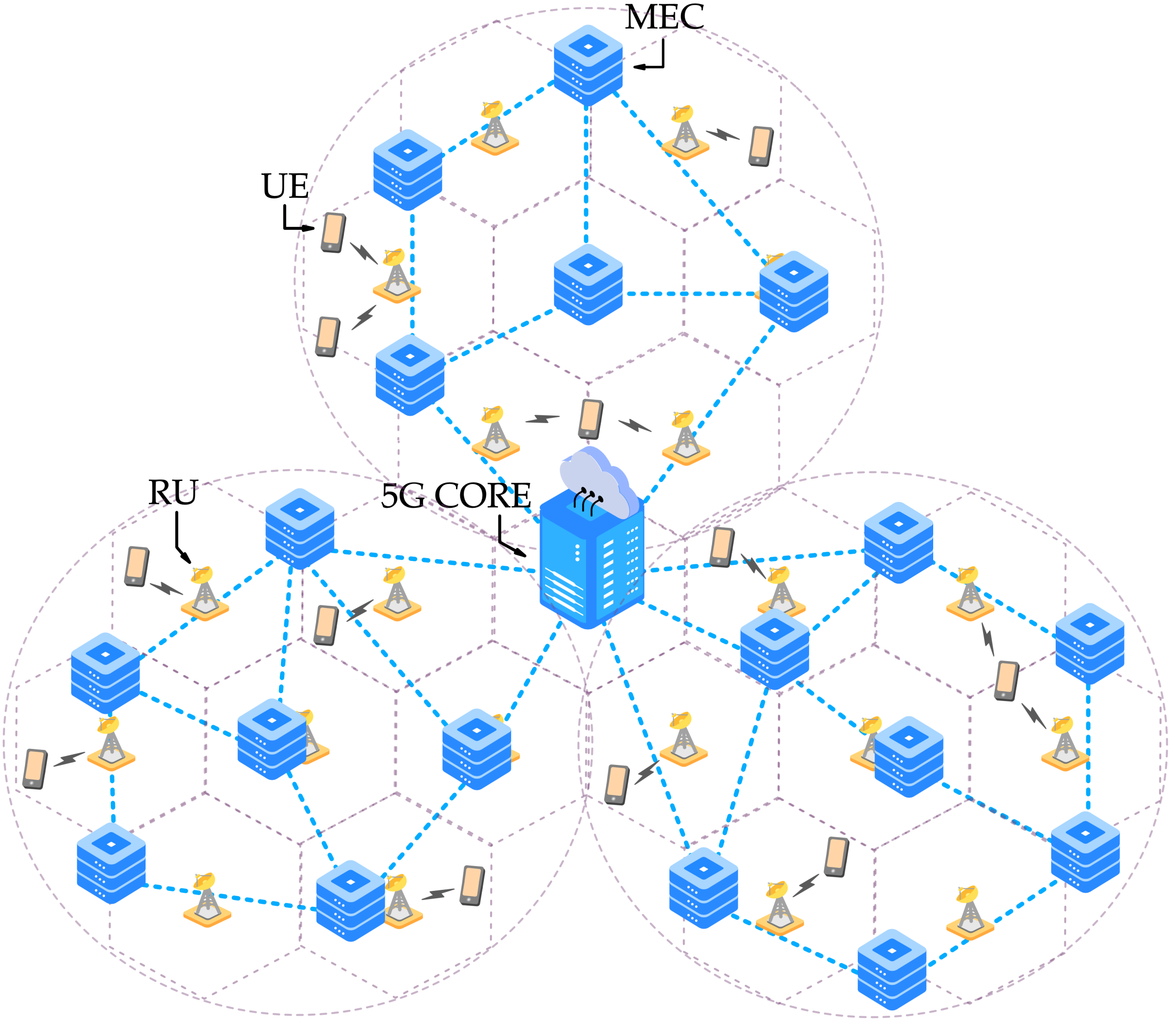}
    \caption{O-RAN in edge computing networks. The network is divided into multiple sub-networks with various network structures based on MEC distributions.}
    \label{scenario}
    \vspace{-0.5cm}
\end{figure}

We take the most representative O-RAN architecture~\cite{wang2022edge} as the basis of the baseband function placement discussion\footnote{This problem formulation is also applicable to other RAN architectures such as C-RAN and Virtualized-RAN (vRAN).}. Without loss of generality, an O-RAN example with edge computing networks is shown in Figure~\ref{scenario}, wherein the MEC networks offer a crucial platform for O-function deployment. 
Affected by the number of service requests in the city, MECs distributed in different districts can be clustered into various sub-networks with different structures~\cite{tranos2015mobile}.
The architecture of a typical network primarily consists of User Equipment (UEs), RU, MECs, and the 5GC. Originating from the UEs, requests of types uRLLC, eMBB, and mMTC will be sent to RUs through the wireless channel. These RUs then aggregate all received requests within their coverage and forward them to baseband function chains, encompassing DU, CU and UPF virtualized on MEC networks, to further access the network service.
These requests vary in their latency and computing resource demands on baseband function chains. Specifically, uRLLC enforces stringent requirements for both fronthaul and end-to-end latency. eMBB predominantly leans on the CU-UP to accommodate its high bandwidth data transmission demands. On the other hand, mMTC, mainly associated with IoT use cases, necessitates greater resources from the CU-CP to handle extensive user access.
Additionally, DU requires dedicated hardware to meet the precise timing synchronization outlined by IEEE 1588~\cite{papatheofanous2021ldpc}. Consequently, certain MECs, lacking the necessary accelerators, are unable to provide DU services.

The objective of \NAME is to devise a baseband function placement strategy to comply with service quality objectives, optimize resource usage, and enhance network power cost-effectiveness. Given that, the baseband function placement problem $P$ can be formulated as follows:
\begin{subequations}\label{eq:subeqns}
 \begin{align}
 \max &\sum_i^I (U_i / ( \sum_n^{N_i} f(\sum_r^{R_i} \sum_v^{V} d_{nrv} c_{rv} / C_n^\prime) + \sum_r^{R_i} \sum_l^{L_i} d_{rl\psi} u_s)) \label{1a} \\
  \text{s.t. \  } &C1: \textstyle \sum_l^{L_i} d_{rl\alpha} p_{l} \leq y_{rf}, \ \forall r \in {R_i}, \ \forall i \in {I} \label{1b} \\
   &C2: \textstyle \sum_l^{L_i} d_{rl\delta} p_{l} \leq y_{rm}, \ \forall r \in {R_i}, \ \forall i \in {I} \label{1c} \\
   &C3: \textstyle \sum_l^{L_i} d_{rl\psi} p_{l} \leq y_{re}, \ \forall r \in {R_i}, \ \forall i \in {I}
 \label{1d}\\
   &C4: \textstyle \sum_r^{R_i} \sum_v^V d_{nrv} c_{rv} \leq C_{n}^\prime,  \ \forall n \in N_i, \ \forall i \in {I} \label{1e}\\ 
   &C5: \textstyle \sum_r^{R_i} \sum_v^V  d_{rl\psi} b_{rv} \leq B_{l}^\prime,  \ \forall l \in {L_i}, \ \forall i \in {I} \label{1f}
 \end{align}
\end{subequations}

For the objective function, the set $I$ signifies the overarching network, comprising multiple sub-networks $i$. Within each sub-network $i$, the sets $N_i$, $R_i$, and $L_i$ represent nodes, requests and links, respectively.
In addition, $U_i$ signifies the overall minimum power consumption in sub-network $i$, which normalizes the objective function. Function $f(*)$, describing the relationship between power cost and resource utilization rate of the MEC server, is measured as per the study in \cite{rahmani2018complete}. $d_{nrv}$ indicates the function placement decision and equals 1 if the baseband function $v$ for request $r$ is positioned on node $n$. $c_{nrv}$ represents the computing resource requirement for the function $v$ associated with request $r$.
$C_n^\prime$ denotes the total computing resource of $n$.
$d_{rl\psi}$ reflects the end-to-end routing decision and equals 1 when traffic for request $r$ heading to its distributed UPF $\psi$ traverses link $l$. The set of baseband functions is denoted as $V$. Within this set, aside from $\psi$, $\alpha$, $\beta$ and $\delta$ correspond to DU, CU-UP and CU-CP, respectively. $u_{s}$ denotes the unified switching power cost. 
Overall, by maximizing the ratio of minimal to actual power consumption, the objective function can be interpreted as minimizing the power consumption across multiple sub-networks by optimizing policies of both the baseband function placement $d_{nrv}$ and routing provisioning $d_{rlv}$. 

Besides, the objective function is subject to the constraints of $C1$ to $C5$. In specific, $C1$, $C2$ and $C3$ ensure the latency requirements of the requests over the network are satisfied, where $p_l$ denotes the length of path $l$, and $y_{rf}$, $y_{rm}$ and $y_{re}$ specify the fronthaul delay limit, midhaul delay limit and end-to-end delay limit of request $r$, respectively. 
Similar to $d_{rl\psi}$ aforementioned, $d_{rl\alpha}$ and  $d_{rl\delta}$ denote fronthaul and midhaul routing decisions. They are set to 1 if traffic for request $r$ to its distributed DU $\alpha$ and CU-CP $\delta$ passes through link $l$, respectively. 
In addition. $C4$ guarantees that the computing resources assigned to baseband functions on each node do not exceed $C_{n}^\prime$. 
The final constraint, $C5$, denotes the capacity limitation of links, with $B_l^\prime$ as the bandwidth capacity of $l$ and $b_{rv}$ as the bandwidth requirement before traffic approaches the baseband function $v$.
The traffic size decreases along the chain, where $b_{rv^\prime} = \epsilon b_{rv^\prime + 1}$ \cite{ndao2023optimal} and $\epsilon$ denotes the traffic decreasing ratio.

Given the binary property of $d_{nrv}$ and $d_{rlv}$ and non-linear function $f(*)$, $P$ is a non-convex combinatorial nonlinear optimization problem. The solution space expands exponentially with increasing network complexity and contains multiple local optima. This makes the problem computationally challenging to address using conventional methods \cite{boyd2004convex}. Therefore, we employ a DRL-based approach, recognized for its effectiveness in managing complex optimization problems\cite{li2022rlops}.


\section{\NAME: Graph DQN-based adaptive baseband function chain placement}
\label{sec:Methodology} 
To infuse intelligence into the network, this paper proposes \NAME. Firstly, the function placement process is transformed into a maze-solving scenario based on MDP, \textit{subject to constraints $C1$ to $C5$}. Subsequently, a DQN-based approach with a \textit{objective function-based reward evaluator} is proposed for its resolution. However, the presence of sub-networks heterogeneity, characterized by distinct $N_i$, $R_i$ and $L_i$ configurations, necessitates multiple DRL training sessions with different neural networks. To overcome this challenge, we introduce a novel feature extraction scheme, which employs a GCN-based encoder and decoder to normalize network features, facilitating an adaptive DRL model for different networks. \NAME, including the Maze-solving environment, GCN-based encoder and decoder and DRL-based solution for solving problem $P$ is discussed in this section.

\begin{figure}[b]
    \vspace{-0.5cm}
    \centering
    \setlength{\abovecaptionskip}{0.05cm}
    \includegraphics[width=0.9\linewidth]{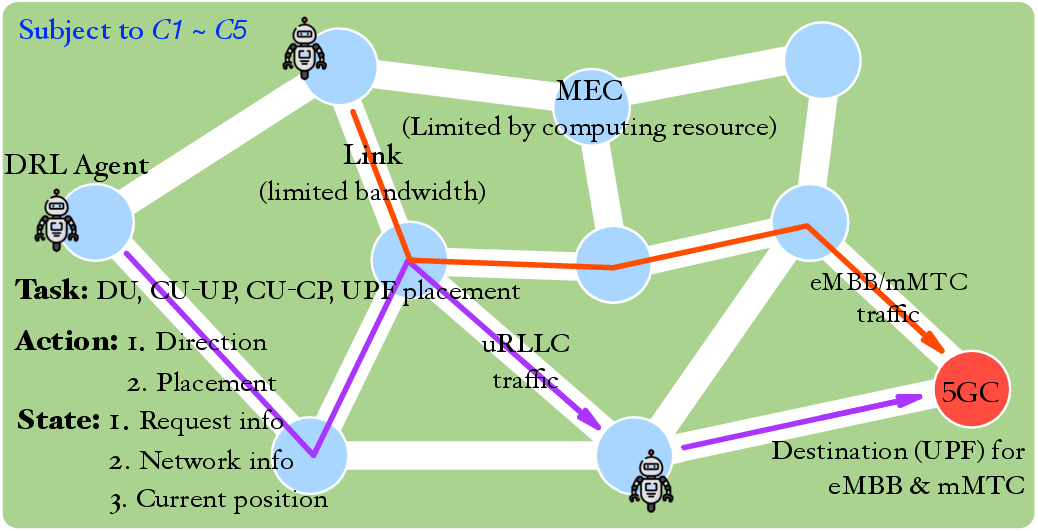}
    \caption{Maze solving scenario.}
    \label{maze}
\end{figure}

\vspace{-0.1cm}
\subsection{Maze-solving Environment}
\label{environment}
The conceptual map of a MEC network as a maze is illustrated in Figure~\ref{maze}.
In this figure, each MEC serves as a node, links between MECs act as paths, and the 5GC stands as the fixed destination for eMBB and mMTC requests. Given the ultra-low latency requirements of uRLLC, its destination can be any node that meets the end-to-end latency constraints.

The maze-solving process strictly follows the sequential-decision making and is formalized as an MDP, which is defined by the tuple $<E_i, A, T, Q, \gamma>$. ${E_i}$ is the set of environment states on sub-network $i$, ${A}$ is the set of actions that an agent performs, ${T}$ represents the transition probability from any state $e\in {E_i}$ to any state $e^\prime \in {E_i}$ for any given action $a \in {A}$.
${Q}$ is the reward function that indicates the immediate reward received from the transition from $e$ to $e^\prime$, and ${\gamma}\in [0,1)$ is the discount factor that trades off the instantaneous and future rewards. The key elements of this MDP are summarized below:

\textit{State $E_i$:} The state information consists of remaining computing resources $C_{ink}$, bandwidth resources $B_{ilk}$ over the network $i$, the current position $Z_{ik}$ in the maze, path lengths $P_i$, and MEC connection relationships $X_i$, as well as the request information $G_{r}$ including $b_{rv}$, $y_{rf}, y_{rm}, y_{re}$, $c_{rv}$ and service type $j_r$ including uRLLC, eMBB, mMTC. Therefore:
\begin{equation}
e_{ik}= \{C_{ink}, B_{ilk}, Z_{ik}, P_i, X_i, G_{r}\} \ \forall n \in N_{i}, \forall r \in R_{i}, \forall i \in I
\end{equation}
where $k$ represents the step account in MDP, and $C_{ink} \leq C_{in}^\prime$, $B_{ilk} \leq B_{il}^\prime$. 

\textit{Action $A$:} The optimization of $d_{nrv}$ and $d_{rlv}$ in problem $P$ is transformed into two actions at each MDP step within the maze-solving context: i) the routing direction $a_{k1}$ and ii) the placement decision at the subsequent node upon arrival $a_{k2}$. 
\begin{equation}
a_{ik}= \{a_{ik1}, a_{ik2}\} \ \ \forall i \in I
\end{equation}
where $a_{ik1} \in [0, m_{i}] \cap \mathbb{Z}$ and $m_{i}$ signifies the maximum out-degree of sub-network $i$. If $a_{ik1} = 0$, the agent stays at the current node. $a_{ik2}$ is a binary number and represents the function placement decision.
The agent executes an action per MDP step, and once the placement for a request is finalized, the agent processes the next request, entering the maze from a different entrance. In the end, an episode is deemed complete after all the requests have been processed.

\textit{Reward $Q$:} The reward is structured in a step-wise manner to motivate the agent to complete the maze. $q_{ik}$ is set to:
\begin{equation}
q_{i k}=\left\{\begin{array}{ll}
-0.2 & \text { if a service fail } \\
0 & \text { if no O-function is placed } \\
0.2 & \text { if an O-function is placed } \\
0.4 & \text { if a service success } \\
g_{i k} & \text { episode ends }
\end{array}\right.
\label{reward}
\end{equation}
The service for a request will be considered as a failure if action $a_{ik}$ is invalid or function placement violates the X-Haul latency constraints, $C1$ to $C3$, computational resource limitations on nodes, $C4$, or bandwidth constraints on links, $C5$. The corresponding reward is set to negative as a penalty. On the other hand, consistent with the objective function of formula~(\ref{1a}), the reward $g_{ik}$ at the end of an episode end is defined as:
\begin{equation}
\begin{aligned}
& g_{i k}= \\
& \left\{\begin{array}{l}
-0.5, \qquad  \  \ \text { if } t<1 / 2 \\
\vspace{0.1cm}
t, \qquad \qquad  \ \text { if } 1 / 2 \leq t<1 \\
\frac{U_{i}}{\displaystyle\sum_{n}^{N_i} f(\displaystyle\sum_{r}^{R_i} \displaystyle\sum_{v}^{V} \frac{d_{nrv} c_{rv}}{C_{n}^\prime} ) + \displaystyle\sum_{r}^{R_i} \displaystyle\sum_{l}^{L_i} d_{rl\psi} u_s} + 1, \ \text{ if } t=1
\end{array}\right.
\end{aligned}
\end{equation}
where $t$ is the ratio of successfully to failed serviced requests in the previous episode. Minimum power consumption $U_{i}$ for sub-network $i$ is achieved by a dynamic slicing window that records $\textstyle \sum_n^{N_{i}} f(\sum_r^{R_i} \sum_v^V d_{nrv} c_{nrv}/C_n^\prime) + \sum_r^{R_i} \sum_l^{L_i} d_{rl\psi} u_s$ in previous 20 episodes in $i$ where $t = 1$.

The maze-based representation of the network can help achieve three objectives: (i) to keep a reasonable action space size $m_i$, (ii) to facilitate the model's generalization, and (iii) to form an MDP where DRL is theoretically applicable.

\subsection{NetMind for Baseband Function Placement}
DQN is adopted to address this maze-solving problem\footnote{The DQN methodology has been well-explained in a plethora of literature, which is omitted in this paper due to the page limits. Readers could refer to~\cite{burhanuddin2023intere} for a better understanding.}. However, due to the variable state lengths, $E_i$, across heterogeneous sub-networks (mazes), distinct DQN models are required for each scenario. That significantly increases the power consumption of this solution on a large-scale deployment.

\begin{figure}[t]
    \centering
    \setlength{\abovecaptionskip}{0.1cm}
    \includegraphics[width=0.85\linewidth]{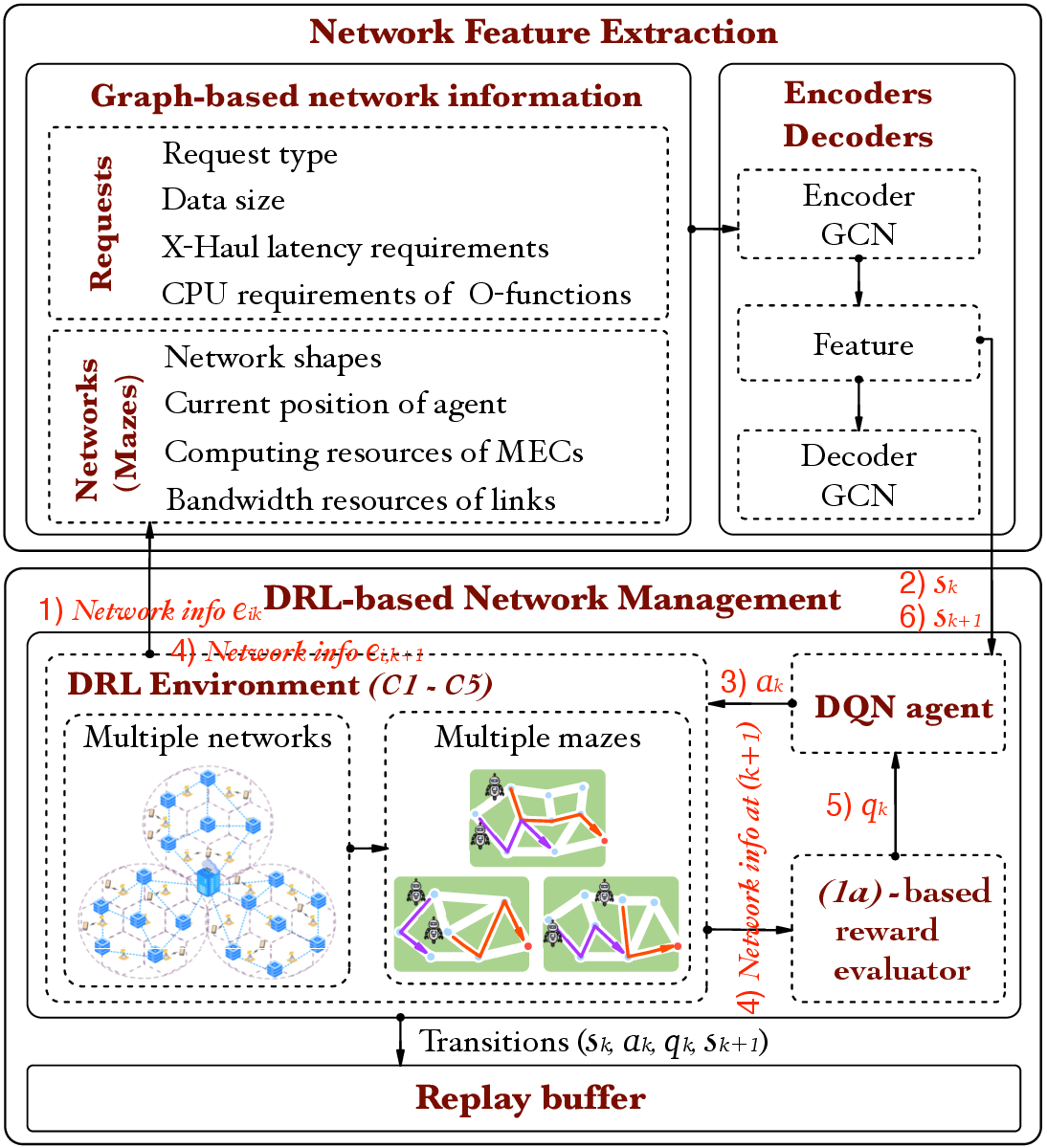}
    \caption{NetMind composition and mechanism.}
    \label{methodology}
    \vspace{-0.5cm}
\end{figure}

To resolve this issue, a novel GCN-based encoder scheme is introduced in \NAME, which processes diverse mazes by converting their different state information, $E_i$, into a uniform-length feature $S$. This feature is used to train one general DRL model fit to different scenarios. Specifically, $\{C_{ink}, Z_{ik}, G_{r}\}$, $\{B_{ilk}, P_i\}$ and $\{X_i\}$ are converted into node attributes, edge attributes, and edge index, respectively. 
In addition, $m_i$ of all sub-networks is also set to a fixed parameter $m_I$, which equals the maximum out-degree over all sub-networks.
A GCN is subsequently employed as the encoder, which aggregates the attributes of each node along with their neighboring nodes. This aggregation is described by:
\begin{equation}
\mathcal{J}_n^{(h)}=\sigma\left(W^{(h)} \sum_{n^\prime \in \mathcal{N}(n) \cup\{n\}} \frac{\mathcal{J}_{n^\prime}^{(h-1)}}{\sqrt{|\mathcal{N}(n)||\mathcal{N}(n^\prime)|}}\right)
\end{equation}
where $\mathcal{J}_n^{(h)}$ represents the feature embedding of node $n$ at the $h^{th}$ layer. $\sigma$ refers to an activation function. $W^{(h)}$ is a trainable weight matrix for the $h^{th}$ layer, which transforms the aggregated feature from the neighboring nodes. $\mathcal{N}(n)$ denotes the immediate neighborhood of node $n$. $\mathcal{J}_{n^\prime}^{(h-1)}$ is the feature embedding of node $n^\prime$ from the previous layer $(h-1)$. The summation over $n^\prime \in \mathcal{N}(n) \cup{n}$ indicates that the features are aggregated not only from the neighboring nodes but also from the node $u$ itself. The term $\sqrt{|\mathcal{N}(n)||\mathcal{N}(n^\prime)|}$ serves as a normalization factor to ensure stable learning. Subsequent to the encoder, a GCN-based decoder is incorporated to interpret these features back to $E_i$, thereby, ensuring the accuracy of the intermediate feature representations.

NetMind generates a new MDP tuple $<{S, A, T, Q, \gamma}>$ and enables \textit{a single, streamlined DQN model} to handle maze-solving problems across different mazes. Instead of training or tuning a new DQN mode, new GCN encoders are required when expanding the complexity of the maze, which saves significant training costs. The overall architecture of \NAME is visualized in Figure~\ref{methodology}.

\section{System Modelling and Simulation}
\label{sec:simulation}
\begin{figure}[t]
    \centering
    \setlength{\abovecaptionskip}{0.1cm}
    \includegraphics[width=0.78\linewidth]{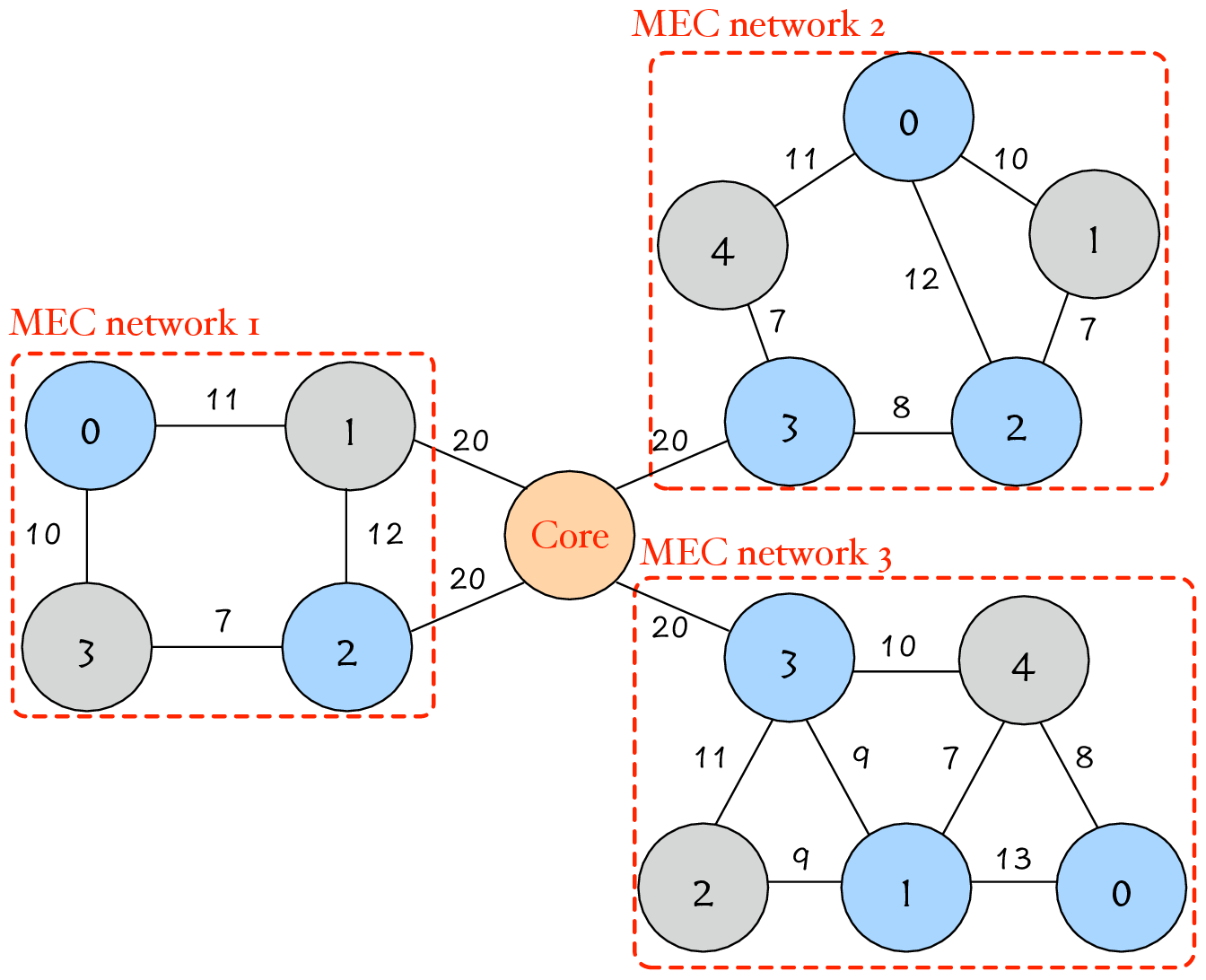}
    \caption{Simulation network consists of three sub-networks with a shared 5GC and various network structures.}
    \label{simulation_network}
    \vspace{-0.5cm}
\end{figure}

\begin{table}[b]
  \centering
  \vspace{-0.5cm}
  \caption{Simulation parameter setup}
  \scalebox{0.95}{
    \begin{tabular}{clrlr}
    \toprule
    \multicolumn{1}{c}{\multirow{4}[2]{*}{Network}} & Node resources $C_n$ (core) &       & 70-100 &  \\
          & Bandwidth $B_{l}$ (Gbps) &       & 30-35 &  \\
          & Switching cost $u_s$ (kW) &       & 30 &  \\
          & Traffic decreasing ratio $\epsilon$ &       & 0.2   &  \\
          & Request amount per episode &       & 0-$N_i$   &  \\
    \midrule
    \multicolumn{1}{c}{\multirow{9}[4]{*}{Request}} &       & \multicolumn{1}{c}{uRLLC} & \multicolumn{1}{l}{eMBB} & \multicolumn{1}{l}{mMTC} \\
\cmidrule{3-5}    & Fronthaul latency $y_{rf}$ (km) & \multicolumn{1}{l}{13-20} & 15-25 & \multicolumn{1}{l}{15-25} \\
          & Midhaul latency $y_{rm}$ (km) & \multicolumn{1}{l}{30-40} & 30-40 & \multicolumn{1}{l}{30-40} \\
          & End-to-end latency $y_{re}$ (km) & \multicolumn{1}{l}{30-40} & 100   & \multicolumn{1}{l}{100} \\
          & DU demands $c_{r\alpha}$ (core) & \multicolumn{1}{l}{15-25} & 15-25 & \multicolumn{1}{l}{15-25} \\
          & CU-UP demands $c_{r\beta}$ (core) & \multicolumn{1}{l}{5-10} & 15-25 & \multicolumn{1}{l}{5-10} \\
          & CU-CP demands $c_{r\delta}$ (core) & \multicolumn{1}{l}{5-10} & 5-10 & \multicolumn{1}{l}{15-25} \\
          & UPF demands $c_{r\psi}$ (core) & \multicolumn{1}{l}{15-25} & 15-25 & \multicolumn{1}{l}{15-25} \\
          & Bandwidth demands $b_{r}$ (Gbps) & \multicolumn{1}{l}{2-4} & 4-7  & \multicolumn{1}{l}{4-7} \\
    \midrule
    \multirow{4}[2]{*}{DRL} & Batch size &       & 50    &  \\
          & Learning rate &       & 0.0002 &  \\
          & Target update coefficient &       & 0.1   &  \\
          & Discount factor &       & 0.99  &  \\
    \midrule
    \multirow{3}[2]{*}{GCN} & Batch size &       & 100   &  \\
          & Learning rate &       & 0.0001 &  \\
          & Extracted feature size &       & 32    &  \\
    \bottomrule
    \end{tabular}%
  }
  \label{setup_table}%
\end{table}%

\subsection{Network Modelling}
The simulated network with three sub-networks is shown in Figure~\ref{simulation_network}. Only the blue nodes are equipped with DU accelerators. Adhering to previous works and the X-Haul latency requirements proposed by IEEE Standards Association~\cite{xiao2021energy, alam2018xhaul}, a setup example is summarized in Table~\ref{setup_table}.

\subsection{Training of NetMind Components}
\begin{figure}[t] 
    \centering 
    \setlength{\abovecaptionskip}{-0.1cm}
    \subfigure[]{
        \label{GCN_train}
        \includegraphics[width=0.46\linewidth]{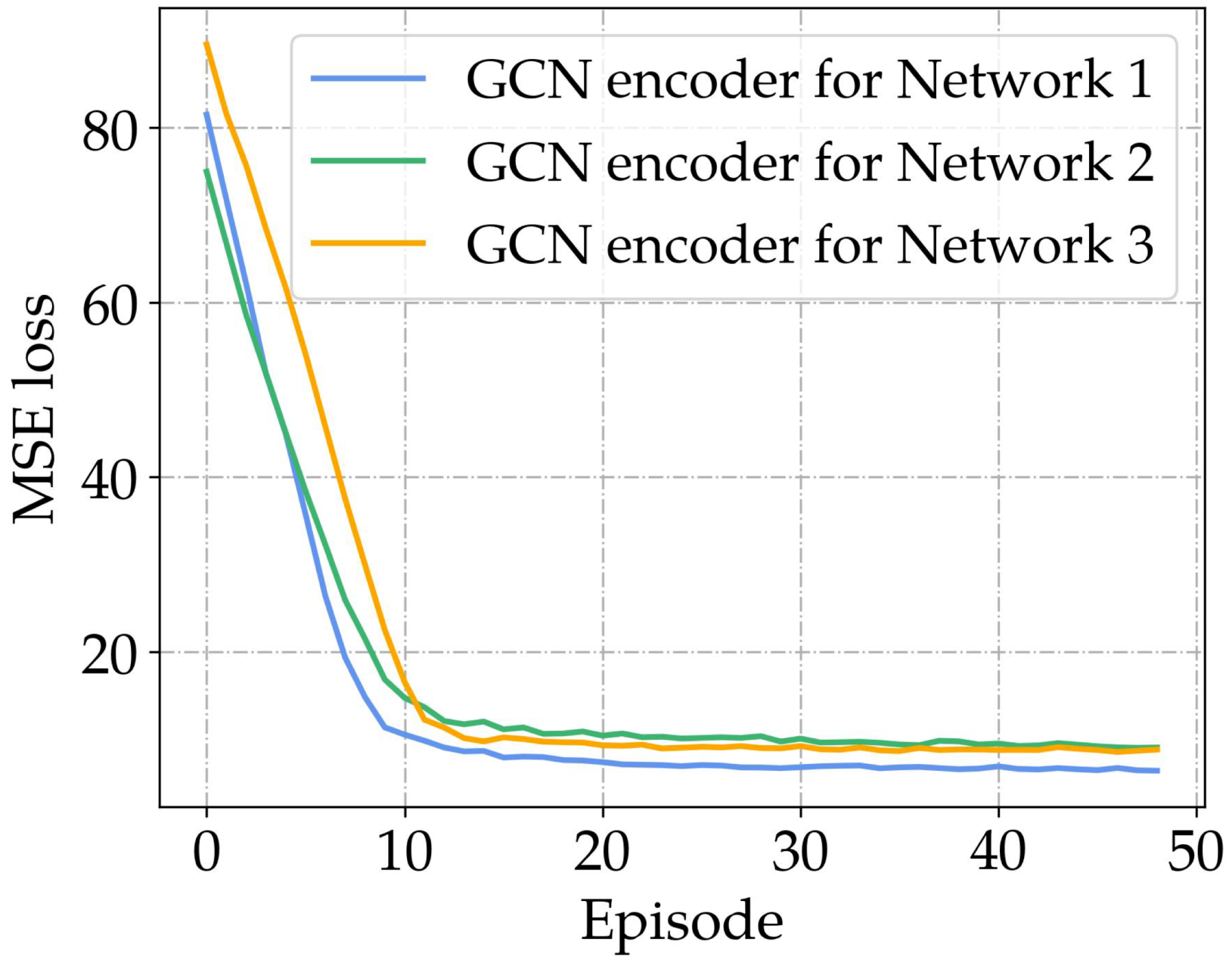}}
    \subfigure[]{
        \label{GDQN_train}
        \includegraphics[width=0.475\linewidth]{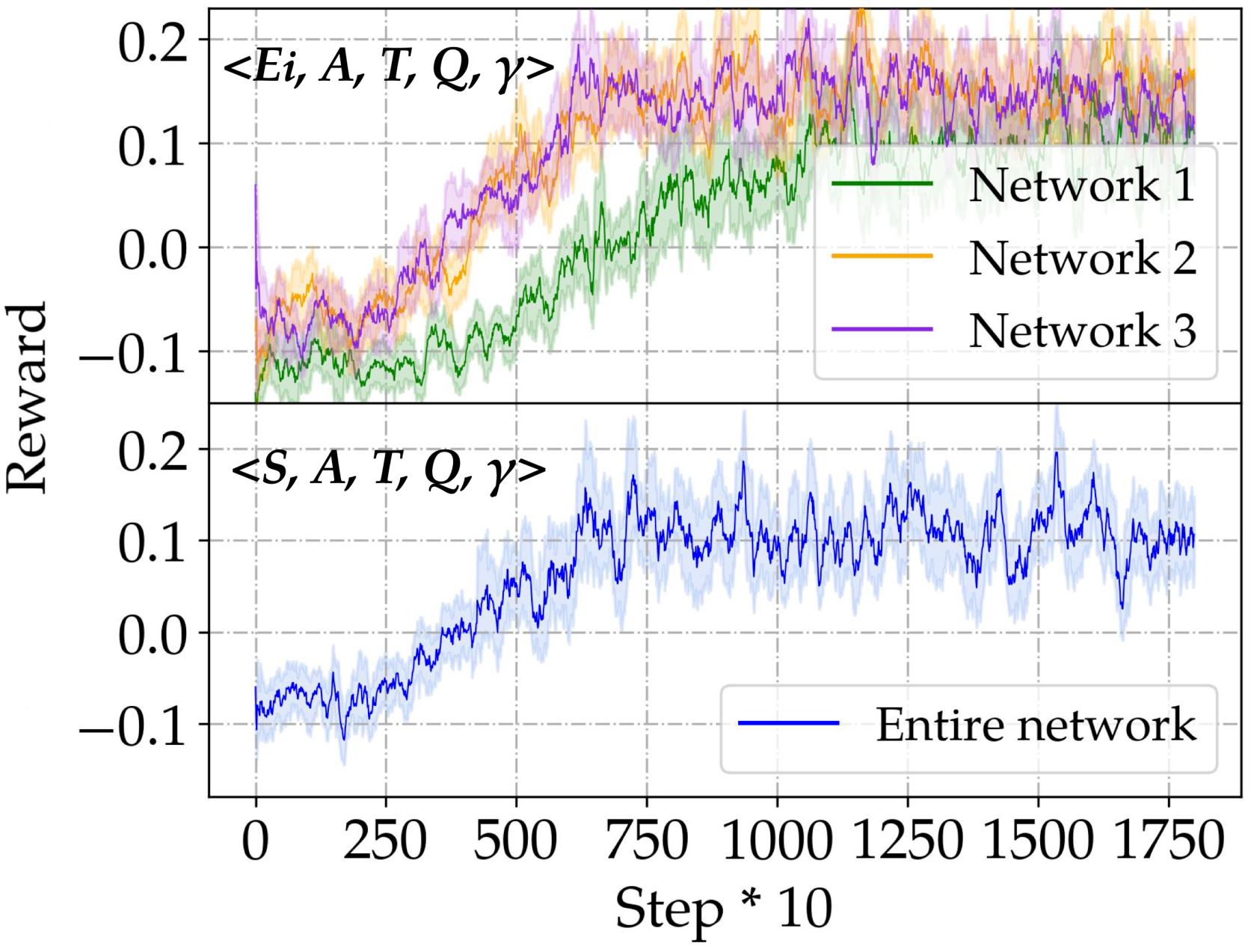}}
    \caption{(a) GCN encoder training performance for three sub-networks; (b) DQN training convergence for different scenarios. Notice the 95-percentile confidence intervals marked over the mean performance curves.} \label{fig:two_trains}
    \vspace{-5pt}
\end{figure}

The training process of the GCN-based encoder and decoder is shown in Figure~\ref{GCN_train}. It ensures the reliability of the extracted features, and more importantly, guarantees the effectiveness of DQN when utilizing these features as state information.
Compared to DQN, which requires frequent interactions with the environment and incurs more communication overhead, the GCN encoder offers higher learning efficiency as a straightforward supervised learning method. To prove that, under the experiment, the average training time of GCN encoders and decoders for sub-networks in Figure~\ref{simulation_network} is evaluated and is less than 40s. 
Therefore, while the \NAME solution necessitates a GCN encoder-decoder for each sub-network, the associated cost is negligible compared to training a new DQN. 

In the case of three sub-networks in Figure~\ref{simulation_network}, Figure~\ref{GDQN_train} demonstrates the training process of \NAME for each sub-network with individual $E_i$ and for the entire network with combined $S$. 
The learning rates of the three circumstances are the same.
As can be seen in Figure~\ref{GDQN_train}, under similar training performance, the training cost of DQN based on $S$ is less than 1/3 of the total training cost for each sub-network trained individually. The proportional relationship between training time and energy consumption can be supported both theoretically and empirically by~\cite{strubell2019energy, garcia2019estimation}. Therefore, the reduction in training time can lead to a significant decrease in energy consumption. 

\vspace{-0.2cm}
\subsection{Numerical Result}
\begin{figure}[t]
    \centering
    \setlength{\abovecaptionskip}{-0.1cm}
    \includegraphics[width=0.78\linewidth]{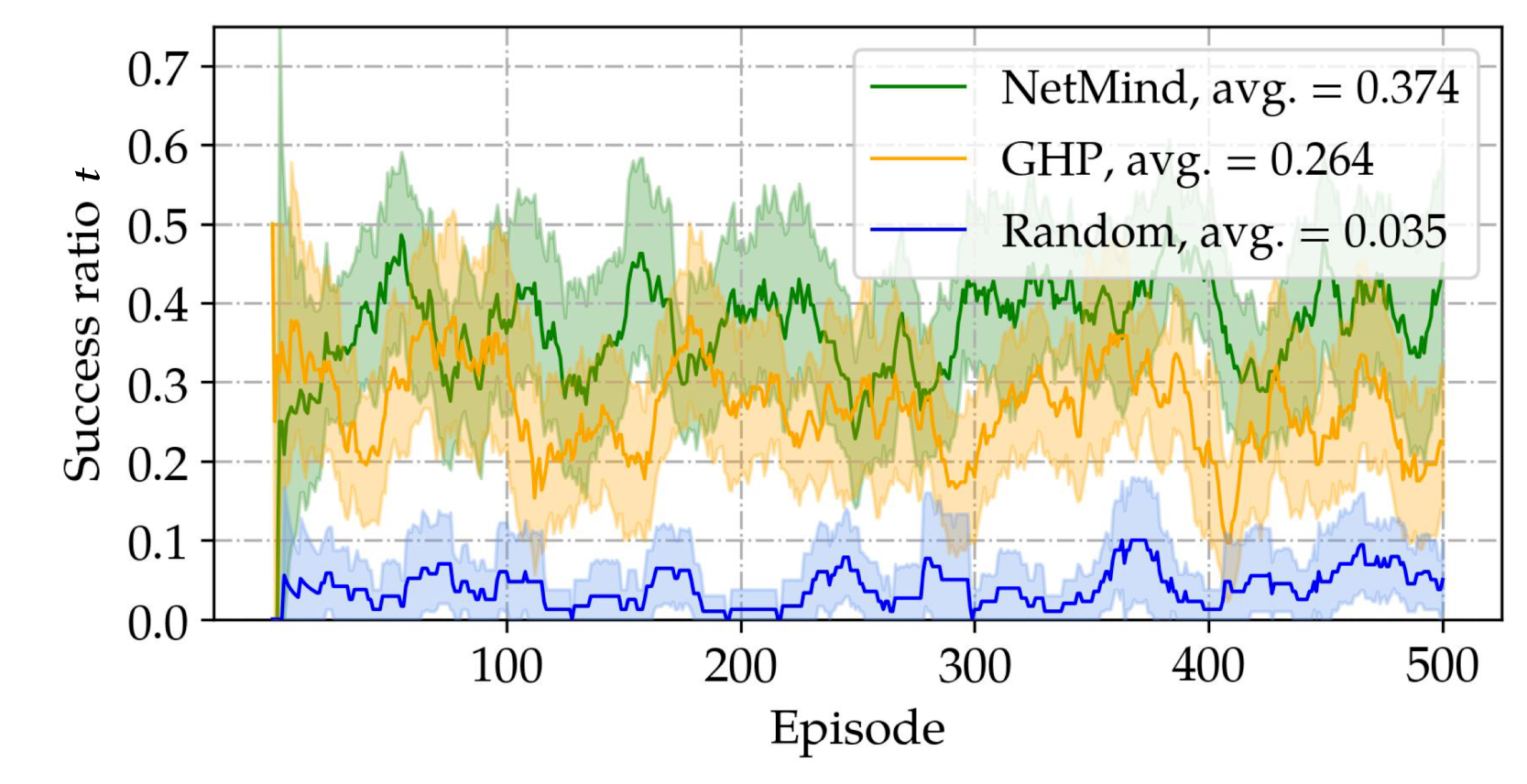}
    \caption{Success ratio $t$ of various baseband function placement strategies. (95-percentile confidence intervals are marked).}
    \label{comparison_1}
    \vspace{-0.5cm}
\end{figure}

\begin{figure}[t] 
    \centering 
    \setlength{\abovecaptionskip}{0cm}
    \subfigure[]{
        \label{comparison_2_1}
        \includegraphics[width=0.46\linewidth]{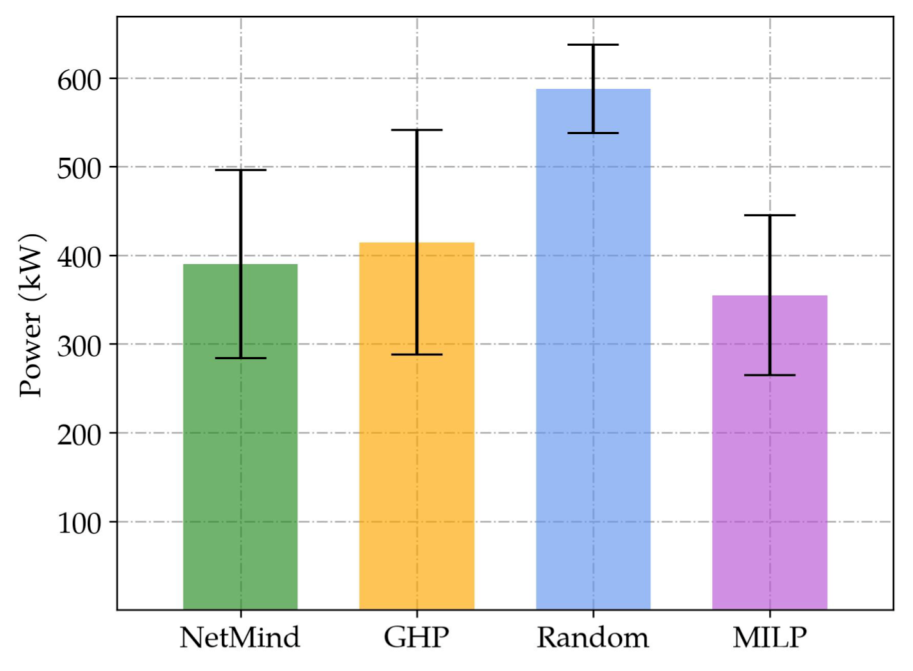}}
    \subfigure[]{
        \label{comparison_2_2}
        \includegraphics[width=0.46\linewidth]{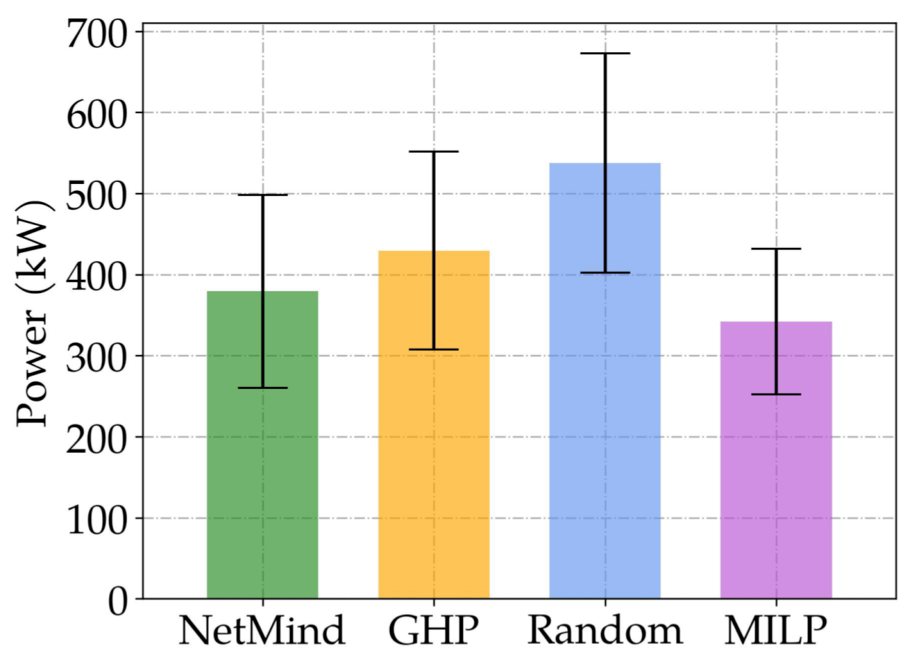}}
    \caption{Mean and standard deviation of power costs for various solutions under (a) \textit{Split option 7.2} based on Table~\ref{setup_table}; and (b) \textit{Split option 6} \cite{rodriguez2020cloud} with a more relaxed fronthaul latency constraint and higher bandwidth requirements.} \label{comparison_2}
    \vspace{-15pt}
\end{figure}

To evaluate the performance of \NAME, as shown in Figure~\ref{comparison_1}, we first explore its success service ratio $t$, which reveals the service stability. Furthermore, Figure~\ref{comparison_2} illustrates the average power consumption per request among three sub-networks over 50 MDP episodes where $t = 1$. This analysis is designed to eliminate the disparities arising from request types, network shapes, and the volume of requests within each $k$. A random allocation solution used in \cite{li2022energy}, a greedy heuristic procedure (GHP) designed by \cite{casazza2017securing}, and a MILP-optimization-based scheme proposed in \cite{xiao2021energy} are adopted to serve as benchmarks. MILP is not assessed in Figure~\ref{comparison_1} due to its time-consuming characteristic.

As can be seen in Figure~\ref{comparison_1} and \ref{comparison_2_1}, NetMind approaches the power-saving performance of the MILP strategy. It surpasses the other two methods, achieving 5\% and 32.76\% superior power-savings and enhancing service stability by 41.67\% and tenfold compared to GHP and the random allocation, respectively. 
Furthermore, to prove the scalability of NetMind in diverse RAN architectures, Figure~\ref{comparison_2_2} demonstrates that it maintains superior performance under another split option (Option 6: characterized by a more relaxed fronthaul latency constraint and higher bandwidth requirements \cite{rodriguez2020cloud}).

\subsection{Discussion and Future Works}
\label{subsec:discussion}
\vspace{-2pt}
\boldpar{Generalization ability} Lack of this ability is a common bottleneck of neural networks due to the input feature distribution drift. In \NAME, the network feature extraction and DRL training are disaggregated into two separate processes, which improves the generalization ability of the DRL agent, as more encoders with diverse features can be incorporated into the DRL training. However, when the network configuration changed significantly, the trained policy still had the risk of being invalid.

\boldpar{Scalability} The scalability of \NAME is twofold: (i) the expandable placement policy, where the GCN encoder adeptly handles the heterogeneity in different O-RAN MEC networks, facilitating the deployment of the placement policy. (ii) the \NAME framework can be applied to other 3GPP splitting options and RAN architectures, with fine-tuning of the parameters and constraints in formula~(\ref{eq:subeqns}). 

\boldpar{Complexity} After deployment, the inference capability of \NAME can approach real-time levels. The training of \NAME consumes the most time, and its complexity is not able to be quantified. As an empirical value from our experiment, in a MacOS with 3.2 GHz 6 CPU cores, it took around 30 minutes to train a DRL model from scratch. This is well acceptable compared with the benchmark MILP having a complexity of $O(N_i^4)$.

\boldpar{Real-world implementation} For \NAME, it is expected to achieve real-time control for the network resources, so the near real-time radio intelligent controller (Near-RT RIC) of O-RAN is the optimal deployment venue. The trained DRL policy can be encapsulated as a microservice application onboarding to Near-RT RIC.

\boldpar{Future Works} 
Regarding future work, we will (i) explore the incremental learning scheme in the DRL algorithm to handle the dynamic change of the network architecture; and (ii) quantify the impact of network feature distribution drift on the effectiveness of DRL policy.

\section{Conclusion}
\label{sec:conclusion}
This paper proposes \NAME, which studies the baseband function and UPF joint placement issue under the context of advanced RAN. This determination of the optimal placement strategy is framed as the DRL-based maze-solving problem. Meanwhile, a novel GCN encoder is introduced as the unified measure of network information extraction to improve the adaptivity of the trained DRL agent in diverse environments. In the simulation, NetMind demonstrates its advantage in reducing training costs and also achieves a 32.76\% power saving over random allocation and a 41.67\% service stability improvement compared to GHP.

\vspace{0cm}
\section*{Acknowledgments}
\small{The authors would like to express their gratitude for the support from the UK-funded projects REASON and TUDOR under the Future Open Networks Research Challenge sponsored by DSIT.}

\bibliographystyle{IEEEtran} %
\scriptsize{\bibliography{IEEEabrv,references}}

\begin{thebibliography}{10}
\providecommand{\url}[1]{#1}
\csname url@samestyle\endcsname
\providecommand{\newblock}{\relax}
\providecommand{\bibinfo}[2]{#2}
\providecommand{\BIBentrySTDinterwordspacing}{\spaceskip=0pt\relax}
\providecommand{\BIBentryALTinterwordstretchfactor}{4}
\providecommand{\BIBentryALTinterwordspacing}{\spaceskip=\fontdimen2\font plus
\BIBentryALTinterwordstretchfactor\fontdimen3\font minus \fontdimen4\font\relax}
\providecommand{\BIBforeignlanguage}[2]{{%
\expandafter\ifx\csname l@#1\endcsname\relax
\typeout{** WARNING: IEEEtran.bst: No hyphenation pattern has been}%
\typeout{** loaded for the language `#1'. Using the pattern for}%
\typeout{** the default language instead.}%
\else
\language=\csname l@#1\endcsname
\fi
#2}}
\providecommand{\BIBdecl}{\relax}
\BIBdecl

\bibitem{3gpp38801}
{3GPP}, ``Study on new radio access technology: Radio access architecture and interfaces,'' 3GPP, Technical Report 38.801, 2016, release 14.

\bibitem{bhamare2018efficient}
D.~Bhamare \emph{et~al.}, ``Efficient virtual network function placement strategies for cloud radio access networks,'' \emph{Comput. Commun.}, vol. 127, 2018.

\bibitem{rodriguez2020cloud}
V.~Q. Rodriguez \emph{et~al.}, ``{Cloud-RAN} functional split for an efficient fronthaul network,'' in \emph{Proc. of the IEEE IWCMC Conf.}, 2020.

\bibitem{yu2020cu}
H.~Yu \emph{et~al.}, ``{DU/CU} placement for {C-RAN} over optical metro-aggregation networks,'' in \emph{Optical Network Design and Modeling: 23rd IFIP WG 6.10 International Conference}.\hskip 1em plus 0.5em minus 0.4em\relax Springer, 2020.

\bibitem{harutyunyan2019latency}
D.~Harutyunyan \emph{et~al.}, ``Latency-aware service function chain placement in {5G} mobile networks,'' in \emph{Proc. of the IEEE NetSoft Conf.}, 2019.

\bibitem{harutyunyan2020latency}
------, ``Latency and mobility--aware service function chain placement in {5G} networks,'' \emph{IEEE TMC}, vol.~21, no.~5, 2020.

\bibitem{zorello2022power}
L.~M.~M. Zorello \emph{et~al.}, ``Power-efficient baseband-function placement in latency-constrained {5G} metro access,'' \emph{IEEE Trans. Green Commun. Netw.}, vol.~6, no.~3, 2022.

\bibitem{joda2022deep}
R.~Joda \emph{et~al.}, ``Deep reinforcement learning-based joint user association and {CU--DU} placement in {O-RAN},'' \emph{IEEE TNSM}, vol.~19, no.~4, 2022.

\bibitem{wang2022edge}
R.~Wang \emph{et~al.}, ``Edge-enhanced graph neural network for {DU-CU} placement and lightpath provision in {X-Haul} networks,'' \emph{JOCN}, vol.~14, no.~10, 2022.

\bibitem{mollahasani2021dynamic}
S.~Mollahasani, M.~Erol-Kantarci, and R.~Wilson, ``Dynamic {CU-DU} selection for resource allocation in {O-RAN} using actor-critic learning,'' in \emph{Proc. of the IEEE GLOBECOM Conf.}, 2021.

\bibitem{li2022energy}
H.~Li \emph{et~al.}, ``{DRL}-based energy-efficient baseband function deployments for service-oriented {Open RAN},'' \emph{IEEE Trans. Green Commun. Netw.}, 2023.

\bibitem{casazza2017securing}
M.~Casazza \emph{et~al.}, ``Securing virtual network function placement with high availability guarantees,'' in \emph{2017 IFIP Networking Conference (IFIP Networking) and Workshops}, 2017.

\bibitem{xiao2021energy}
Y.~Xiao, J.~Zhang, and Y.~Ji, ``Energy-efficient {DU-CU} deployment and lightpath provisioning for service-oriented {5G} metro access/aggregation networks,'' \emph{J. Light. Technol.}, vol.~39, no.~17, 2021.

\bibitem{tranos2015mobile}
E.~Tranos and P.~Nijkamp, ``Mobile phone usage in complex urban systems: a space--time, aggregated human activity study,'' \emph{J. Geogr. Syst.}, vol.~17, no.~2, 2015.

\bibitem{papatheofanous2021ldpc}
E.~A. Papatheofanous, D.~Reisis, and K.~Nikitopoulos, ``The {LDPC} challenge in software-based 5g new radio physical layer processing,'' in \emph{Proc. of the IEEE MeditCom Conf.}, 2021.

\bibitem{rahmani2018complete}
R.~Rahmani, I.~Moser, and M.~Seyedmahmoudian, ``A complete model for modular simulation of data centre power load,'' \emph{arXiv:1804.00703}, 2018.

\bibitem{ndao2023optimal}
A.~Ndao \emph{et~al.}, ``Optimal placement of virtualized dus in {O-RAN} architecture,'' in \emph{Proc. of the IEEE VTC2023-Spring Conf.}, 2023.

\bibitem{boyd2004convex}
S.~Boyd, S.~P. Boyd, and L.~Vandenberghe, \emph{Convex optimization}.\hskip 1em plus 0.5em minus 0.4em\relax Cambridge university press, 2004.

\bibitem{li2022rlops}
P.~Li \emph{et~al.}, ``Rlops: Development life-cycle of reinforcement learning aided open ran,'' \emph{IEEE Access}, vol.~10, 2022.

\bibitem{burhanuddin2023intere}
L.~A.~B. Burhanuddin \emph{et~al.}, ``Inter-cell interference mitigation for cellular-connected {UAVs} using {MOSDS-DQN},'' \emph{IEEE TCCN}, 2023.

\bibitem{alam2018xhaul}
\BIBentryALTinterwordspacing
Alam. (2018, 3) Dimensioning challenges of xhaul. IEEE Task Force. [Online]. Available: \url{https://sagroups.ieee.org/1914/wp-content/uploads/sites/92/2018/03/tf1_1803_Alam_xhaul-dimensioning-challenges_1.pdf}
\BIBentrySTDinterwordspacing

\bibitem{strubell2019energy}
E.~Strubell, A.~Ganesh, and A.~McCallum, ``Energy and policy considerations for deep learning in {NLP},'' \emph{arXiv:1906.02243}, 2019.

\bibitem{garcia2019estimation}
E.~Garc{\'\i}a-Mart{\'\i}n \emph{et~al.}, ``Estimation of energy consumption in machine learning,'' \emph{JPDC}, vol. 134, 2019.

\end{thebibliography}

\end{document}